\documentstyle[twoside,fleqn,espcrc2,epsfig]{article}

\def\slashchar#1{\setbox0=\hbox{$#1$}           
   \dimen0=\wd0                                 
   \setbox1=\hbox{/} \dimen1=\wd1               
   \ifdim\dimen0>\dimen1                        
      \rlap{\hbox to \dimen0{\hfil/\hfil}}      
      #1                                        
   \else                                        
      \rlap{\hbox to \dimen1{\hfil$#1$\hfil}}   
      /                                         
   \fi}                                         %

\newcommand{\ifig}[1]{\mbox{\epsfig{file=#1,height=50mm,width=75mm}}}

\def\bc{\begin{center}}
\def\ec{\end{center}}
\def\be{\begin{equation}}
\def\ee{\end{equation}}
\newcommand{\chilim}{\lim_{m \rightarrow 0}}
\newcommand{\ba}{\begin{eqnarray}}  
\newcommand{\ea}{\end{eqnarray}}

\newcommand{\nn}{\nonumber}

\newcommand{\bea}{\begin{eqnarray}}
\newcommand{\eea}{\end{eqnarray}}

\newcommand{\as}{\alpha_s} 

\newcommand{\W}{\hat{   W}}
\newcommand{\U}{\hat{ U}}
\newcommand{\gammaz}{\hat{\gamma}^{(0)T}}

\newcommand{\AmS}{{\protect\the\textfont2
  A\kern-.1667em\lower.5ex\hbox{M}\kern-.125emS}}

\title{Weak Matrix Elements without Quark Masses on the Lattice}

\author{L.~Giusti\address{Department of Physics, Boston University
         Boston, MA 02215 USA.}\thanks{Invited talk at the QCD Euroconference
         99. Work done in collaboration with A.~Donini, V.~Gimen\'ez and
         G.~Martinelli. This research was supported in part under DOE 
         grant DE-FG02-91ER40676.}}
       
\begin{document}

\begin{abstract}
We introduce  a new parameterization of four-fermion  matrix elements which
does not involve quark masses and thus allows a reduction of   
systematic uncertainties in physical amplitudes. As a result the 
apparent quadratic dependence of $\epsilon'/\epsilon$ on $m_s(\mu)$ is 
removed. To simplify the matching
between  lattice and continuum renormalization schemes, we
express our results in terms of   Renormalization Group Invariant
$B$-parameters which  are renormalization-scheme and scale 
independent. As an application of our proposal, matrix elements 
of  $\Delta I=3/2$ 
and SUSY $\Delta F =2$ ($F=S,C,B$) four-fermion operators  have been computed.
\end{abstract}

\maketitle

\section{Introduction}
\label{sec:intro}
Since  the original proposals of using lattice QCD to study hadronic 
weak decays~\cite{CMP,BGGM,bernardargo}, 
substantial theoretical and numerical progress has been made:
the main theoretical aspects of the renormalization of composite four-fermion
operators  are well  understood \cite{boc,shape1,4ferm_teo}; the calculation 
of  $K^0$--$\bar K^0$ mixing, relevant to the prediction of the CP-violation
parameter $\epsilon$, has reached a level of accuracy which 
is unpaired by any other approach \cite{Laur};  
increasing precision has been gained in the determination of the
electro-weak penguin amplitudes necessary to  the prediction
of the CP-violation parameter $\epsilon^\prime/\epsilon$ 
\cite{gupta_bp,NoiDELTAS=2,ds2susy}; finally matrix  elements of $\Delta S=2$ 
operators which are relevant to study FCNC effects in SUSY models have been 
computed~\cite{NoiDELTAS=2,ds2susy}. Methods and symbols used in this
talk  and all the results we report are fully described 
in \cite{NoiDELTAS=2,ds2susy}.

\section{Matrix elements without quark masses}
\label{sec:definitions}
The analysis of $K^0-\bar K ^0$ mixing with the most
general $\Delta S =2$ effective Hamiltonian requires   
the knowledge of the matrix elements $\langle\bar K^0|O_i |K^0 \rangle$ of the 
parity conserving parts of the following operators 
\bea 
O_1 &=& \bar s^\alpha \gamma_\mu (1- \gamma_{5} ) d^\alpha \ 
\bar s^\beta \gamma_\mu (1- \gamma_{5} )  d^\beta ,  \nn \\ 
O_2 &=& \bar s^\alpha (1- \gamma_{5} ) d^\alpha \ 
 \bar s^\beta  (1- \gamma_{5} )  d^\beta ,  \nn \\ 
O_3&=& \bar s^\alpha  (1- \gamma_{5} )  d^\beta  \ 
 \bar s^\beta   (1- \gamma_{5} ) d^\alpha ,  \label{eq:ods2} \\ 
O_4 &=& \bar s^\alpha  (1- \gamma_{5} ) d^\alpha \  
\bar s^\beta  (1 + \gamma_{5} )  d^\beta ,  \nn \\ 
O_5&=& \bar s^\alpha  (1- \gamma_{5} )  d^\beta \ 
 \bar s^\beta (1 +  \gamma_{5} ) d^\alpha . \nonumber
 \eea
On the lattice, matrix elements of weak four-fermion operators 
are computed from first principles. But, following the common lore, 
they are usually 
given in terms  of the  so-called $B$-parameters which measure 
the deviation  of their values from those obtained in the Vacuum 
Saturation Approximation (VSA). 
For operators in (\ref{eq:ods2}), the
$B$-parameters are usually defined as
\bea\label{eq:bparC} 
\langle  \bar K^{0} \vert   O_{1} (\mu) \vert K^{0} 
\rangle &=& \frac{8}{3} M_{K}^{2} f_{K}^{2} B_{1}(\mu) \, ,  \\
\langle  \bar K^{0} \vert  O_{i} (\mu) \vert K^{0} \rangle  &=&
\frac{C_i}{3} \left( \frac{ M^2_{K} f_K }{ m_{s}(\mu) + 
m_d(\mu) }\right)^{2} B_{i}(\mu)\, ,\nonumber
\eea
where $C_i=-5,1,6,2$ for ($i=2,\dots , 5$).
In (\ref{eq:bparC}), $\langle \bar K^0|O_1 |K^0 \rangle$ is
parameterized in terms of well-known experimental  quantities and $B_1(\mu)$ 
($B_K(\mu)\equiv B_1(\mu)$). On the contrary, $\langle \bar K^0|O_i |K^0 \rangle$ 
($i=2,\dots ,5$) depend quadratically on the quark masses in (\ref{eq:bparC}),
while they are expected to remain finite in the chiral limit and depend only  
linearly on the quark masses. Contrary to $f_{K}$, $M_{K}$,  etc., 
quark masses  can not be  
directly measured by experiments and the present accuracy  
in their determination is still rather  poor. 
\begin{table}[tbhp]
\setlength{\tabcolsep}{.10pc}
\newlength{\digitwidth} \settowidth{\digitwidth}{\rm 0}
\catcode`?=\active \def?{\kern\digitwidth}
\vspace{-0.6cm}
\caption{\it{Matrix elements in GeV$^4$ at the renormalization scale 
$\mu = 2$~GeV in 
the RI scheme. In the first two columns
the  results obtained with the new 
parameterization are given, while in the last two columns we show the results obtained 
in ref.~\cite{NoiDELTAS=2} with the ``conventional'' parameterization on the 
same set of data.}}
\begin{tabular}{||c|c|cc||c|cc||}\hline\hline
      & \multicolumn{2}{c}{New}& &\multicolumn{2}{c}{Old}& \\
\hline
$\langle  O_i \rangle$ &$\beta=6.0$&$\beta=6.2$& 
& $\beta=6.0$  &$\beta=6.2$& \\
 &this work& this work & &\cite{NoiDELTAS=2}
&\cite{NoiDELTAS=2}& \\
\hline \hline
$\langle  O_1 \rangle$& 0.012(2) & 0.011(3) & & 0.012(2) & 0.011(3) & \\ 
$B_1$   &                   0.70(15) & 0.68(21) & & 0.70(15) & 0.68(21) & \\
\hline
$\langle  O_2 \rangle$&-0.079(10)&-0.074(8) & &-0.073(15)&-0.073(15)& \\ 
$B_2$                     & 0.72(9)  &0.67(7)   & & 0.66(3)  & 0.66(4)  & \\
\hline
$\langle  O_3 \rangle$&0.027(2)  & 0.021(3) & &0.025(5)  &0.022(5)  & \\ 
$B_3$                     &1.21(10)  & 0.95(15) & & 1.12(7)  & 0.98(12) & \\
\hline
$\langle  O_4 \rangle$& 0.151(7) & 0.133(12)& &0.139(28) &0.133(28) & \\ 
$B_4$                     & 1.15(5)  & 1.00(9)  & & 1.05(3)  & 1.01(6)  & \\
\hline
$\langle  O_5 \rangle$& 0.039(3) & 0.029(5) & &0.035(7)  &0.029(7)  & \\ 
$B_5$                     & 0.88(6)  & 0.66(11) & & 0.79(6)  & 0.67(10) & \\
\hline
$\langle  O^{3/2}_7 
\rangle$                  & 0.019(2) & 0.011(3) & & 0.020(5) & 0.014(5) & \\ 
$B^{3/2}_7$               & 0.65(5)  & 0.38(11) & & 0.68(7)  & 0.46(13) & \\
\hline
$\langle  O^{3/2}_8 
\rangle$                  & 0.082(4) & 0.068(8) & & 0.092(19)& 0.087(19)& \\ 
$B^{3/2}_8$               & 0.92(5) &0.77(9) & & 1.04(4) & 0.98(8) & \\
\hline \hline
\end{tabular}
\label{tab:summary}
\vspace{-0.5cm}
\end{table}
Therefore,
whereas for $O_1$ we introduce $B_{K}$ 
as an alias of the matrix element,  by using (\ref{eq:bparC}) we 
replace each of the SUSY matrix elements with 2 unknown quantities, i.e. 
the $B$-parameter and $m_{s} + m_d$.
To overcome these problems, 
we  propose the following new parameterization of $\Delta S =2$ operators 
\bea\label{eq:FURBAdef}
\langle  \bar K^{0} \vert   O_{1} (\mu) \vert K^{0} 
\rangle &=& \frac{8}{3} M_{K}^{2} f_{K}^{2} B_1(\mu) ,  \\
\langle  \bar K^{0} \vert   O_{i} (\mu) \vert K^{0} \rangle  &=&
M_{K^*}^{2} f_{K}^{2} \tilde{B}_i(\mu).\nn
 \eea
The $\tilde B_{i}(\mu)$ parameters are still dimensionless
quantities and can be computed on the lattice by studying
appropriate ratios of three- and two point functions \cite{ds2susy}.
\begin{table}[htb]
\centering
\vspace{-0.4cm}
\caption{\it{RGI Matrix elements in GeV$^4$ computed as in Eq.~(\ref{eq:BRGI}) 
with $\alpha_s^{n_f=4}$.}}
\label{tab:summaryrgi}
\begin{tabular}{||c|c|c||}\hline\hline
$\langle  O^{RGI}_i\rangle$ &$\beta=6.0$&$\beta=6.2$\\
\hline \hline
$\langle  O^{RGI}_1 \rangle$ & 0.017(3) & 0.016(4) \\ 
\hline
$\langle  O^{RGI}_2 \rangle$ & -0.051(7) & -0.048(6) \\ 
\hline
$\langle  O^{RGI}_3 \rangle$ & 0.005(7) & -0.004(7) \\ 
\hline
$\langle  O^{RGI}_4 \rangle$ & 0.072(3) & 0.063(6) \\ 
\hline
$\langle  O^{RGI}_5 \rangle$ & 0.043(3) & 0.032(5) \\ 
\hline \hline
\end{tabular}
\end{table}
By simply using them, we have eliminated any fictitious reference 
to the quark masses, hence reducing the systematic errors on the 
corresponding physical amplitudes. An alternative parameterization 
which has not been used in our numerical analysis,  
but may be very useful in the future, can be found in \cite{ds2susy}.\\
The VSA and $B$-parameters are also used for matrix elements of  
operators which enter the $\Delta S =1$ effective Hamiltonian.
Notice that 
this "conventional" parameterization is the only responsible for the apparent 
quadratic dependence of $\epsilon'/\epsilon$ on the quark masses.
This introduces a redundant source of systematic error which can be 
avoided by parameterizing the matrix elements in terms of measured
experimental quantities and 
therefore {\it a better determination of the strange quark mass 
$m_s(\mu)$ will not improve our theoretical knowledge of 
$\epsilon'/\epsilon$}. 
In this work we have computed the matrix
elements $\langle\pi|O^{3/2}_{i}|K\rangle$ of the four fermion operators
$O^{3/2}_{i}$ ($i=7,8,9$) which contribute to the $\Delta I=3/2$ sector of 
$\epsilon'/\epsilon$. In fact  
in the chiral limit 
$\langle\pi\pi|O^{3/2}_{i}|K\rangle$  
can be obtained,  using  soft pion theorems,  from  
$\langle\pi^{+}|O^{3/2}_{i}|K^{+}\rangle$. For degenerate quark 
masses, $m_{s}=m_{d}=m$, and in the chiral limit,  we find
\ba\label{eq:softpion}
\chilim \langle\pi^{+}|O^{3/2}_{7}|K^{+}\rangle & = & 
 - M_{\rho}^{2} f_{\pi}^{2} \chilim \tilde{B}_5(\mu)\nonumber\\
\chilim \langle\pi^{+}|O^{3/2}_{8}|K^{+}\rangle & = &
 - M_{\rho}^{2} f_{\pi}^{2} \chilim \tilde{B}_4(\mu)\nonumber\\
\chilim \langle\pi^{+}|O^{3/2}_{9}|K^{+}\rangle & = &
\frac{8}{3} M_{\pi}^{2} f_{\pi}^{2} \chilim B_1(\mu)\; .  \nonumber
\ea
In the limit $m_s=m_d$ complicated subtractions of lower dimensional
operators are avoided for $\Delta I=3/2$ operators. This is not the case for
$\Delta I=1/2$ operators which enter the determination of 
$\epsilon'/\epsilon$ : in this case the mixing with lower dimensional
operators makes the computation much more involved. A reliable lattice
estimate of these matrix elements is still missing but encouraging
preliminary results with domain-wall fermions have been presented in 
ref.~\cite{Blum}.
    
\section{Renormalization Group Invariant Operators}
\label{sec:RGI}
Physical amplitudes can be written as 
\be 
\langle F \vert {\cal H}_{eff}\vert I \rangle = 
\langle F \vert \vec{O}(\mu) \vert I \rangle \cdot \vec{C}(\mu) 
\, , \label{wope} 
\ee
where $\vec{ O}(\mu) \equiv
( O_1(\mu),\dots,  O_N(\mu))$ is the operator basis (for example the basis  defined 
in (\ref{eq:ods2}) for the $\Delta S = 2$ effective Hamiltonian) and
$\vec{C}(\mu)$   the corresponding Wilson coefficients
represented as a column vector.
$\vec C(\mu)$ is  expressed in terms of its counter-part, 
computed at a large scale $M$, 
through the renormalization-group evolution matrix  $\W[\mu,M]$
\be 
\vec C(\mu) = \W[\mu,M] \vec C(M)\, , \label{evo} 
\ee
where the initial conditions $\vec C(M)$, are obtained by 
perturbative matching 
of the full theory to the effective one 
at the scale $M$ where all the heavy particles have been removed. 
$\W[\mu,M]$ can be 
written as (see for example \cite{scimemi})
\bea
\W[\mu,M]  = \hat M[\mu] \U[\mu, M] \hat M^{-1}[M] \, , 
 \label{monster} \eea
where $\U = (\alpha_s(M)/\alpha_s(\mu))^{(\gamma^{(0)T}_O/2\beta_0)}$ is the 
leading-order evolution matrix
and $M(\mu)$ is a NLO matrix defined in \cite{scimemi} which can 
be obtained by solving 
the Renormalization Group Equations (RGE) at the next-to-leading order.
The Wilson coefficients $\vec{C}(\mu)$ and the  renormalized operators $\vec{ O}(\mu)$ 
are usually defined  in a given scheme, at a fixed renormalization scale $\mu$, 
and they depend on the renormalization scheme and scale in such a way 
that only $H_{eff}$ is scheme and scale independent. This is a source of confusion in 
the literature, especially when (perturbative) coefficients and  
(non-perturbative) matrix elements are computed using different techniques, regularization, 
schemes and renormalization scales.
To simplify the matching procedure, we propose a Renormalization
Group Invariant (RGI) definition of Wilson coefficients and 
composite operators 
which generalizes what is usually done for $B_K$ 
and for quark masses~\cite{martiRGI,luscherRGI}. We define 
\be\label{eq:CRGI}
\hat{w}^{-1}[\mu] \equiv  \hat M[\mu] \left[\as (\mu) \right]^{
      - \gammaz_O / 2\beta_{ 0}}\, , 
\ee
and using  Eqs.~(\ref{monster}) and (\ref{eq:CRGI}) we obtain
\be
\W[\mu,M]  = \hat{w}^{-1}[\mu]\hat{w}[M]\; .  
\ee
The effective Hamiltonian (\ref{wope}) can be written as 
\ba\label{eq:HRGIbella}
{\cal H}_{eff} 
    =     \vec O^{RGI} \cdot \vec C^{RGI}\; ,  
\ea 
where
\be\label{eq:monster2}
\vec C^{RGI}      =  \hat{w}[M] \vec C(M)\; , \;
\vec O^{RGI}     =   \vec{ O}(\mu) \cdot \hat{w}^{-1}[\mu]\; .
\ee
{\it $\vec C^{RGI}$ and  $\vec O^{RGI}$ are scheme and scale independent at  
the order we are working. Therefore
the effective Hamiltonian is splitted in terms which are individually 
scheme and scale independent}. This procedure is 
generalizable to any effective weak Hamiltonian.
The $\tilde{B}$-parameters defined in eqs.~(\ref{eq:FURBAdef}) satisfy 
the same RGE as the corresponding 
operators and the RGI $\tilde{B}$-parameters can be defined as 
\be\label{eq:BRGI}
{\tilde B}_i^{RGI} = \sum_{j} \tilde{B}_j(\mu)   w(\mu)_{ji}^{-1}\, .  
\ee
\begin{figure}[tbh]
\ifig{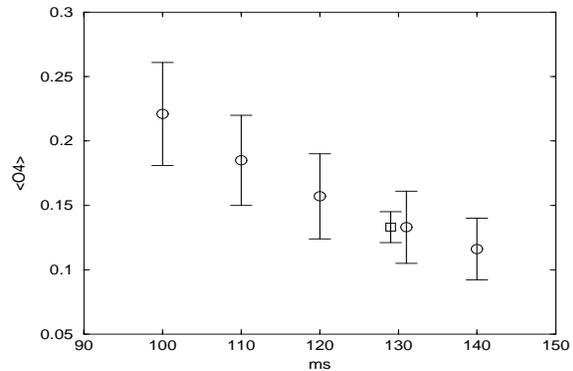}
\vspace{-1.1cm}
\caption{\small{Comparison of 
$\langle  \bar K^{0} \vert O_{4} \vert K^{0} \rangle$ in GeV$^4$  
computed with the new (square) and the old (circle) parameterization as a 
function of the strange quark mass used in (\ref{eq:bparC}) to obtain the
full matrix element from the $B$ parameters.}}
\label{fig:fig.1}
\vspace{-0.5cm}
\end{figure} 

\section{Numerical results}
\label{sec:numeri}
All details concerning the extraction of matrix elements from 
correlation functions and the computation of 
the non-perturbative renormalization constants of  
lattice operators can be found in \cite{4ferm_teo,NoiDELTAS=2,ds2susy}. 
In this talk we report the results obtained in \cite{ds2susy}. 
The simulations have  been performed at $\beta = 6.0$ 
(460 configurations) and   $6.2$  (200 configurations) 
with the tree-level Clover
action, for several values of the quark masses and for different 
meson momenta. The physical volume is 
approximatively the same on the two lattices. Statistical errors have been 
estimated with the jacknife method. The main results we have 
obtained for $\Delta S =2$ and $\Delta I=3/2$ matrix elements and their
comparison with the results in \cite{NoiDELTAS=2} are 
reported in Tables~\ref{tab:summary} and \ref{tab:summaryrgi}.
\begin{figure}[tbh]
\ifig{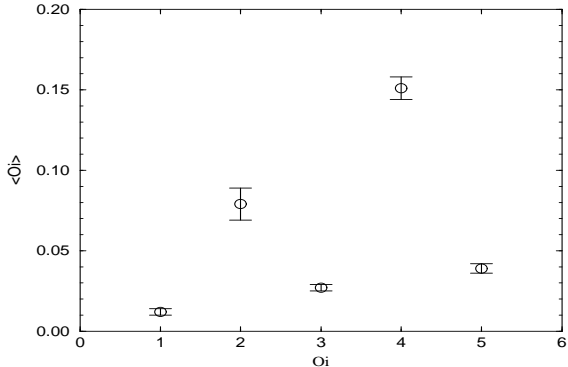}
\vspace{-1.1cm}
\caption{\small{Values of 
$\langle  \bar K^{0} \vert O_{i} \vert K^{0} \rangle$ in GeV$^4$ computed at 
$\mu=2$~GeV in the RI scheme. $O_1$ corresponds to the Standard Model 
$\Delta S =2$ operator.}}
\label{fig:fig.2}
\vspace{-0.5cm}
\end{figure} 
In Figure~\ref{fig:fig.1} 
we show the strong dependence of 
$\langle  \bar K^{0} \vert O_{4} \vert K^{0} \rangle$ 
on the strange quark mass when the "conventional"
parameterization (\ref{eq:bparC}) is used, to be compared 
with the results obtained with the new parameterization. 
It is also evident that with the
same set of data the new parameterization allows to determine the  
matrix elements with smaller systematic uncertainties.   
Although  we have data at two different values of the lattice spacing, the 
statistical errors, and the uncertainties in the extraction of the 
matrix elements,  are too large to enable any extrapolation 
to the continuum limit $a \to 0$ : within the precision of our results  
we cannot study the dependence of $\tilde B$-parameters on $a$. 
 \begin{table}[htb]
\centering
\caption{\it{Matrix elements in GeV$^4$ at $\mu=2$~GeV in the RI scheme and 
their RGI values with $\alpha_s^{n_f=4}$.}}
\label{tab:summarybest}
\begin{tabular}{||c|c|c||}\hline\hline
$\langle  O_i\rangle$ & RI & RGI \\
\hline \hline
$\langle  O_1 \rangle$ & 0.012(3)   & 0.017(4) \\ 
\hline
$\langle  O_2 \rangle$ & -0.077(10) & 0.050(7)\\ 
\hline
$\langle  O_3 \rangle$ & 0.024(3)   & 0.001(7)\\ 
\hline
$\langle  O_4 \rangle$ & 0.142(12)  & 0.068(6)\\ 
\hline
$\langle  O_5 \rangle$ & 0.034(5)   & 0.038(5)\\ 
\hline \hline
\end{tabular}
\vspace{-0.5cm}
\end{table}
For this reason, we estimate our best values of the $B$-parameters
by  averaging the results obtained at the two values of 
$\beta$ \cite{ds2susy}. Since the results at $\beta=6.0$ have smaller
statistical errors but suffer from larger discretization effects, we do not
weight the averages with the quoted statistical errors but take simply the
sum of the two values divided by two. As far as the errors are concerned, 
we take the largest of the two statistical errors. Our best results are
reported in Table~\ref{tab:summarybest} and are shown in 
fig.~\ref{fig:fig.2}. It is interesting to note, as expected from chiral 
perturbation theory, that matrix elements of $\Delta S =2$ SUSY 
operators are enhanced respect to the SM one ($\langle \hat O_1 \rangle$) 
by a factor $2-12$ at $\mu=2$~GeV. Therefore, low energy QCD effects can 
enhance contributions  beyond the Standard Model to $\epsilon_K$ 
\cite{jhep,strumioski} which, compared with the other SM predictions, 
becomes a promising observable to detect signals of new physics at low 
energy.
The results for the analogous $\Delta C=2$ and $\Delta B =2$ matrix elements 
are reported in \cite{NOIDELTAB=2}.

\end{document}